\title{World-line perturbation theory}

\author{Jan-Willem van Holten\footnote{Email: v.holten@nikhef.nl} \\
        Nikhef \\
        Science Park 105 \\
        1098 XG Amsterdam, Netherlands}

\date{Oct. 31, 2016}

\documentclass[11pt,english]{article}

\usepackage[]{graphicx}
\usepackage{amsfonts}

\newcommand{\nit}{\noindent}

\newcommand{\np}{\newpage}
\newcommand{\dsp}{\displaystyle}
\newcommand{\vs}[1]{\vspace{#1 ex}}
\newcommand{\hs}[1]{\hspace{#1 em}}
\newcommand{\bfr}{\begin{flushright}}
\newcommand{\efr}{\end{flushright}}
\newcommand{\bc}{\begin{center}}
\newcommand{\ec}{\end{center}}
\newcommand{\ben}{\begin{enumerate}}
\newcommand{\een}{\end{enumerate}}

\newcommand{\be}{\begin{equation}}
\newcommand{\ee}{\end{equation}}
\newcommand{\ba}{\begin{array}}
\newcommand{\ea}{\end{array}}
\newcommand{\ct}{\cite}

\newcommand{\ag}{\alpha}
\newcommand{\bg}{\beta}
\newcommand{\gam}{\gamma}
\newcommand{\del}{\delta}

\newcommand{\ve}{\varepsilon}
\newcommand{\zg}{\zeta}
\newcommand{\thg}{\theta}
\newcommand{\kg}{\kappa}
\newcommand{\lb}{\lambda}
\newcommand{\sg}{\sigma}
\newcommand{\rg}{\rho}

\newcommand{\vf}{\varphi}
\newcommand{\og}{\omega}
\newcommand{\Gam}{\Gamma}
\newcommand{\Del}{\Delta}

\newcommand{\Sg}{\Sigma}

\newcommand{\brr}{\bar{r}}

\newcommand{\bfer}{\bold{r}}

\newcommand{\bfx}{\bold{x}}

\newcommand{\bfJ}{\bold {J}}

\newcommand{\lh}{\left(}
\newcommand{\rh}{\right)}

\newcommand{\nb}{\nabla}

\newcommand{\ctg}{\mbox{\,cotan\,}}

\newcommand{\der}{\partial}

\begin{document}

\maketitle

\begin{abstract}
The motion of a compact body in space and time is commonly described by the world line 
of a point representing the instantaneous position of the body. In General Relativity such 
a world-line formalism is not quite straightforward because of the strict impossibility to 
accommodate point masses and rigid bodies. In many situations of practical interest it can 
still be made to work using an effective hamiltonian or energy-momentum tensor for a 
finite number of collective degrees of freedom of the compact object. Even so exact 
solutions of the equations of motion are often not available. In such cases families of 
world lines of compact bodies in curved space-times can be constructed by a perturbative 
procedure based on generalized geodesic deviation equations. Examples for simple test 
masses and for spinning test bodies are presented. 
\end{abstract}
 
\nit
\section{Test bodies in General Relativity} 

The newtonian theory of gravity is a theory of instantaneous action at a distance, which is 
consistent with the concept of absolute time and absolute simultaneity. This allows for the 
existence of rigid bodies. Taking that for granted Newton proved in the {\em Principia} that 
the orbital motion of a homogeneous spherical rigid body is correctly represented by that 
of a point mass located at the center of gravity. Thus he was able to explain the motion of 
the moon orbiting the earth. Also in more general situations the motion of a rigid body can 
be represented by a single curve: its world line, identified as the orbit in space and time of 
the center of gravity,  whilst the remaining kinematical degrees of freedom are restricted to 
rotational motion of the body about the center of gravity and specify the orientation of the 
body at every point of the world line. In this frame work there is no obstacle to consider the 
limit of a very small rigid body of which the gravitational influence on the motions of other 
bodies is negligeable. Such a body which probes the gravitational field without disturbing 
it is commonly referred to as a test body. It is characterized by a finite number of physical 
degrees of freedom as its motion is completely specified in terms of its world line plus 
orientational degrees of freedom. 

In General Relativity (GR) the situation is more subtle, as strictly point-like particles cannot
be accommodated in the theory. Most importantly any object of finite non-zero mass $m$ 
has an associated Schwarzschild radius 
\[
\rg_S = \frac{2Gm}{c^2}, 
\]
such that if the size of the object shrinks below this scale it becomes a black hole with 
a finite surface area equal to $A = 4 \pi \rg_S^2$ for spherical bodies. Thus any 
massive body is an extended body, at least in classical GR. In situations where quantum 
effects become relevant this may change, but then there are other limitations like the 
Compton wave length opposing complete localization of objects. 

If a classical object also possesses an internal angular momentum (spin) there are 
additional complications. First of all in GR rigid bodies can not exist and there is no 
unique, observer-independent center of mass. Indeed as there is no absolute 
simultaneity the relative positions of different particles composing the body at any 
fixed time depend on the state of motion of the body with respect to the observer. 
Some elementary considerations showing this state of affairs are discussed for a 
simple two-body system in appendix A. In addition, in a relativistic context for a 
composite system it is more appropriate to discuss the motion of a center of energy 
or {\em centroid} rather than a center of mass, although such a concept is still 
observer-dependent. Actually it has been shown \ct{moller1949, moller1957,f.costa2014} 
that under reasonable assumptions the world lines of all possible centroids of an object 
with mass $m$ and spin $s$ fill a time-like oriented tube of radius 
\[
\rg_M = \frac{s}{mc}.
\]
The upshot of this discussion is that strictly speaking in GR no unique world line 
can be associated with the motion of massive bodies and any particular choice of 
representative world line is at least in part a matter of convenience and requires 
careful specification. 

Nevertheless there are circumstances in which the localization of a compact body 
is possible with sufficient accuracy that for practical purposes it may be regarded 
as a mass point moving along a world line. Moreover if its mass is small enough 
that one can neglect the associated space-time curvature and its influence on other 
bodies, one can still regard such an approximate point mass as a test body probing 
the space-time geometry in which it moves. 

In the context of GR the space-time geometry and the motion of compact bodies 
are linked by the Einstein equations\footnote{Here and in the following we use natural
units in which the speed of light $c = 1$.}
\be
G_{\mu\nu} + 8\pi G\, T_{\mu\nu} = 0,
\label{1v2.1}
\ee
where the Einstein tensor $G_{\mu\nu}$ is specified by the space-time geometry, and 
the energy-momentum tensor $T_{\mu\nu}$ describes the physical degrees of matter.
Irrespective of the precise background geometry the Bianchi identities for the Einstein  
tensor require the energy-momentum tensor to be divergence-free: 
\be
\nb_{\mu} T^{\mu\nu} = 0.
\label{1v2.2}
\ee
For a compact body with mass $m$ moving as an approximate point mass along a 
world line $\xi^{\mu}(\tau)$ parametrized by the proper time $\tau$ the effective 
energy-momentum tensor is \ct{weinberg1972} 
\be 
T^{\mu\nu}(x) = \frac{m}{\sqrt{-g}}\, \int d\tau\, \dot{\xi}^{\mu} \dot{\xi}^{\nu} \del^4(x - \xi(\tau)),
\label{1v2.3}
\ee
with the overdot denoting a derivative w.r.t.\ proper time\footnote{In this paper the delta 
function is defined as a scalar density of weight $1/2$ such that \\ 
$~~~~~\int d^4y\, f(y)\, \del^4(y - x) = f(x)$.}. 
Its divergence vanishes if the world line is a geodesic: 
\be
\nb_{\mu} T^{\mu\nu} = \frac{m}{\sqrt{-g}}\, \int d\tau \left[ \ddot{\xi}^{\nu} + 
 \Gam_{\mu\lb}^{\;\;\;\nu}(\xi) \dot{\xi}^{\mu} \dot{\xi}^{\lb} \right] \del^4(x - \xi(\tau)) = 0.
\label{1v2.4}
\ee
Neglecting the back reaction of the compact body is allowed if the contributions of the 
test body to the geometry of space time are too small to be of interest.  As an illustration 
take the local background geometry to be that of flat Minkowski space-time:
\be
g^{(0)}_{\mu\nu} = \eta_{\mu\nu} \hs{1} \mbox{and} \hs{1} \Gam^{(0)\;\nu}_{\mu\lb} = 0.
\label{1v2.5}
\ee
In particular we can then choose to work in the local inertial frame in which the body is at rest:
\be
T^{(0)}_{00} = m \del^3(\bfx), \hs{1} T^{(0)}_{i0} = T^{(0)}_{ij} = 0 \hs{2} i,j = (1,2,3).
\label{1v2.6}
\ee
Taking into account these energy-momentum source terms, the solution of the 
Einstein equation is modified to first order to read 
\be
g^{(1)}_{\mu\nu} = \eta_{\mu\nu} + h_{\mu\nu}, 
\label{1v2.7}
\ee
where the correction term satisfies the linearized Einstein equation 
\be 
\ba{l} 
\hs{-2} \Box h_{\mu\nu} - \der_{\mu} \der^{\lb} h_{\lb\nu} - \der_{\nu} \der^{\lb} h_{\lb\mu} 
 + \der_{\mu} \der_{\nu} h_{\lb}^{\;\,\lb}  - \eta_{\mu\nu} \lh \Box h_{\lb}^{\;\,\lb}
 - \der^{\kg} \der^{\lb} h_{\kg\lb} \rh \\
 \\
\hs{-2} = - 8\pi G\, T^{(0)}_{\mu\nu}.
\ea 
\label{1v2.8}
\ee 
Removing gauge degress of freedom by the De Donder conditions 
\[
\der^{\mu} h_{\mu\nu} = \frac{1}{2}\, \der_{\nu} h_{\mu} ^{\;\,\mu},
\]
the Einstein equation simplifies further to 
\be
\Box \lh h_{\mu\nu} - \frac{1}{2}\, \eta_{\mu\nu} h_{\lb}^{\;\lb} \rh = - 16\pi G\, T^{(0)}_{\mu\nu},
\label{1v2.9}
\ee
which has the solution
\be
h_{i0} = 0, \hs{1} h_{ij} = \del_{ij} h_{00}, \hs{1} h_{00} = \frac{2Gm}{r}.
\label{1v2.10}
\ee
With this correction we obtain the modified line element 
\be
g^{(1)}_{\mu\nu} dx^{\mu} dx^{\nu} = - \lh 1 - \frac{2 Gm}{r} \rh dt^2 + \lh 1 + \frac{2Gm}{r} \rh d\bfer^2. 
\label{1v2.11}
\ee
It follows that the geometry near the test body deviates strongly from flat space on the scale 
of its Schwarzschild radius, and it can be considered as a near point-like object only as long 
as the external curvature is comparatively small on this scale. 

Actually to first order in $G$ the geometry specified by the line element (\ref{1v2.11}) 
coincides with that of Schwarzschild space-time in isotropic co-ordinates: 
\be
g_{\mu\nu} dx^{\mu} dx^{\nu} = - \lh \frac{2r - Gm}{2r + Gm} \rh^2 dt^2 
  + \lh 1 + \frac{Gm}{2r} \rh^4 d\bfer^2.
\label{1v2.12}
\ee
In hindsight it is not such a surprise that the test-particle approximation is the limit of a black-hole 
geometry, as it is well-known that the standard black-hole space-times are exact vacuum 
solutions of the Einstein equations characterized by a finite number of parameters like 
mass, spin and electric charge, in which respect they resemble test bodies. The difference
is of course that their action on the space-time geometry has been fully taken into 
account to the extent that --in contrast to test particles-- their energy-momentum tensor has 
been absorbed completely in the space-time curvature, i.e.\ it has become part of the Einstein 
tensor in eqn.\ (\ref{1v2.1}).

\section{The motion of test bodies}

Restricting our considerations to situations where the internal degrees of freedom of a 
compact body do not take part in its gravitational interaction and its size and gravitational 
back reaction can be neglected, its motion can be represented adequately by a world line 
which is a time-like geodesic of the external space-time geometry. This is the simplest 
case of the test-body approximation. The more elaborate test-body limit of a compact 
body with spin will be discussed in sects.\ 4 and later. 

This description of a test body as an object with a finite numer of degrees of freedom 
to which one can assign at any moment a representative position has considerable 
mathematical advantage: to such a body we can associate a finite-dimensional phase 
space in which the evolution of the system is described by a simple curve. This orbit is 
generated by a hamiltonian $H$ depending on a finite number of phase-space variables 
including position and momentum. As by construction it neglects the finite size and 
gravitational back reaction of the body, it is of course an {\em effective hamiltonian}, 
its validity restricted by the test-body limit. 

The effective hamiltonian dynamics of a massive test body with no other degrees of 
freedom, like spin or charge, and an energy-momentum tensor of the form (\ref{1v2.1}) 
is straightforward to construct. The phase space is spanned by the position 
variables  $\xi^{\mu}(\tau)$ and the momentum variables $\pi_{\mu}(\tau)$, with the 
usual equal proper-time canonical Poisson brackets 
\be
\left\{ \xi^{\mu}, \pi_{\nu} \right\} = \del_{\nu}^{\mu}.
\label{2.8}
\ee
The free hamiltonian 
\be
H = \frac{1}{2m}\, g^{\mu\nu}[\xi]\, \pi_{\mu} \pi_{\nu}
\label{2.9}
\ee
then generates the equations of motion 
\be
\dot{\xi}^{\mu} = \left\{ \xi^{\mu}, H \right\} = \frac{1}{m}\, g^{\mu\nu} \pi_{\nu}, \hs{2}
\dot{\pi}_{\mu} = \left\{ \pi_{\mu}, H \right\} = \frac{1}{2m}\, \der_{\mu} g^{\nu\lb}\, \pi_{\nu} \pi_{\lb}.
\label{2.10}
\ee
By simple algebra these equations can be rewritten in the form 
\be
\pi_{\mu} = m g_{\mu\nu} \dot{\xi}^{\nu}, \hs{2} 
\ddot{\xi}^{\mu} + \Gam_{\lb\nu}^{\;\;\;\mu} \dot{\xi}^{\lb} \dot{\xi}^{\nu} = 0,
\label{2.10.1}
\ee
reproducing the geodesic equation derived earlier from the energy-momentum 
tensor (\ref{1v2.1}). 

In this hamiltonian frame work it is easy to find constants of motion, simplifying the 
solution of the geodesic equation. There is at least one universal constant independent 
of the specific metric $g_{\mu\nu}$, the hamiltonian itself: 
\be
H = - \frac{m}{2} \hs{1} \Leftrightarrow \hs{1} g_{\mu\nu}\, \dot{\xi}^{\mu} \dot{\xi}^{\nu} = - 1.
\label{2.11}
\ee
It establishes the usual relation between proper time and co-ordinate time. 
Other constants of motion depend on the symmetries of the background space-time, as 
implied by Noether's theorem. For example a quantity 
\be 
J[\xi,\pi] = \ag^{\mu}[\xi]\, \pi_{\mu}
\label{2.12}
\ee
is a constant of motion if $\ag^{\mu}$ is a Killing vector, representing an isometry of the 
metric: 
\be
\dot{J} = \left\{ J, H \right\} = 0 \hs{1} \Leftrightarrow \hs{1} 
\nb_{\mu} \ag_{\nu} + \nb_{\nu} \ag_{\mu} = 0.
\label{2.13}
\ee
As an  example in static space-times the kinetic energy $E$ of the test body is a constant 
of motion: 
\be
\ag^{\mu} = (-1, 0 , 0 , 0) \hs{1} \Rightarrow \hs{1} E = - \pi_t.
\label{2.14}
\ee
In particular in Minksowki space-time:
\be 
E = -m\, \eta_{tt}\, \dot{\xi}^t = m\, \frac{dt}{d\tau} = \gam m, 
\label{2.15}
\ee
where $\gam$ is the time-dilation factor. 

In spherically symmetric  geometries, such as Minkowski, Schwarzschild and 
Freedmann-Lemaitre type cosmological space-times, all three components of 
angular momentum are conserved: 
\be
\ba{l}
J_1 = - \sin \vf\, \pi_{\thg} - \ctg \thg\, \cos \vf\, \pi_{\vf}, \\
 \\
J_2 = \cos \vf\, \pi_{\thg} - \ctg \thg\, \sin \vf\, \pi_{\vf}, \\
 \\
J_3 = \pi_{\vf}.
\ea
\label{2.16}
\ee
Here the momenta $(\pi_r, \pi_{\thg}, \pi_{\vf})$ are defined w.r.t.\ a polar co-ordinate 
frame $(r,\thg,\vf)$ such that the hamiltonian is given by
\be
2m H = - g^{tt}(t,r) \pi_t^2 + g^{rr} (t,r) \pi_r^2 + g^{\thg\thg}(t,r) \lh \pi_{\thg}^2 + 
 \frac{\pi_{\vf}^2}{\sin^2 \thg} \rh.
\label{2.17}
\ee
Clearly if the metric components $g_{\mu\nu}$ are time independent the system 
is both static and spherically symmetric, and energy an angular momentum are all
conserved. This holds in particular for Minkowski and Schwarzschild space-time.

The standard metric of Schwarzschild space-time in Droste co-ordinates for an object
with mass $M$ can be obtained from the isotropic co-ordinate representation (\ref{1v2.12}) 
by the co-ordinate transformation
\be
\brr = r \lh 1 + \frac{GM}{2r} \rh^2,
\label{2.18}
\ee
reproducing the static, spherically symmetric line element
\be
g_{\mu\nu} dx^{\mu} dx^{\nu} = - \lh 1 - \frac{2GM}{\brr} \rh dt^2 
 + \frac{d\brr^2}{1 - \frac{2GM}{\brr} } + \brr^2 \lh d\thg^2 + \sin^2 \thg\, d\vf^2 \rh.
\label{2.19}
\ee
Replacing $\brr \rightarrow r$ we can then write the hamiltonian for a test body in 
Schwarzschild space-time as
\be
2m H = - \frac{\pi_t^2}{1 - \frac{2GM}{r}} + \lh 1 - \frac{2GM}{r} \rh \pi_{r}^2 
 + \frac{1}{r^2} \lh \pi_{\thg}^2 + \frac{\pi_{\vf}^2}{\sin^2 \thg} \rh, 
\label{2.20}
\ee
which is a special instance of a static hamiltonian of the type (\ref{2.17}).
The conservation of kinetic energy then holds in the form 
\be
\ve \equiv \frac{E}{m} = \lh 1 - \frac{2GM}{r} \rh \frac{dt}{d\tau}. 
\label{2.21}
\ee
Rotating the co-ordinate system such that the plane of the orbit is at constant  
$\thg = \pi/2$, we have $J_1 = J_2 = \pi_{\thg} = 0$ and
\be
\ell \equiv \frac{J_3}{m} = r^2\, \frac{d\vf}{d\tau}.
\label{2.22}
\ee
In addition the universal constant of motion $H = - m/2$ implies that 
\be
\lh \frac{dr}{d\tau} \rh^2 = \ve^2 - \lh 1 - \frac{2GM}{r} \rh \lh 1 + \frac{\ell^2}{r^2} \rh.
\label{2.23}
\ee
In particular there are circular orbits $r = R =$ constant for which 
\be
\ba{l}
\dsp{ \frac{d\vf}{d\tau} = \frac{\ell}{R^2} = \sqrt{ \frac{GM}{R^3}}\, \frac{1}{\sqrt{1 - \frac{3GM}{R}}}, }\\
 \\
\dsp{ \frac{dt}{d\tau} = \frac{\ve}{1 - \frac{2GM}{R}} = \frac{1}{\sqrt{1 - \frac{3GM}{R}}}. }
\ea
\label{2.24}
\ee
Observe that these equations reproduce Keplers third law for circular orbits:
\[
\frac{d\vf}{dt} = \sqrt{\frac{GM}{R^3}} \hs{1} \Rightarrow \hs{1} T^2 = \frac{4\pi^2}{GM}\, R^3,
\]
where $T$ is the orbital period measured by the clock of a distant observer keeping  
co-ordinate time $t$, and the radial co-ordinate $R$ is determined by the orbital 
circumference through the relation $L = 2 \pi R$.

\section{Geodesic deviations} 

In most space-time geometries general exact solutions of the geodesic equation are difficult 
to obtain, and when they are available they are often expressed in terms of non-elementary 
transcendental functions \ct{hackmann2015}. However given one particular geodesic curve 
it is possible to find approximate solutions for arbitrary nearby geodesics using the geodesic 
deviation equation and its higher-order generalizations \ct{kerner2001,kerner2002}. This 
procedure can also be explained in terms of the Schild ladder construction 
\ct{schild1972,mtw1970}, for which next-to-leading order corrections can be obtained using 
the results reviewed in this section. 

In terms of the tangent vector $u^{\mu} = \dot{\xi}^{\mu}$ the geodesic equation (\ref{2.10.1}) 
reads
\be
D_{\tau} u^{\mu} = \dot{u}^{\mu} + \Gam_{\lb\nu}^{\;\;\;\mu} u^{\lb} u^{\nu} = 0.
\label{3.0}
\ee
Now consider a family of world lines 
\be
\xi^{\mu}(\tau,\sg) = \xi^{\mu}(\tau) + \sg n^{\mu}(\tau),
\label{3.1}
\ee
obtained by a displacement of the geodesic in the direction of a vector field $n^{\mu}(\tau)$ 
scaled by the real parameter $\sg$. The corresponding changes in the world line and 
its tangent vector are described covariantly by the vectors
\be
\Del \xi^{\mu} = \sg n^{\mu}, \hs{1} \Del u^{\mu} = D_{\tau} \Del \xi^{\nu} = \sg D_{\tau} n^{\mu}.
\label{3.2}
\ee
For this displacement to respect the geodesic equation (\ref{3.0}) to first order in $\sg$ we
require 
\be
\Del D_{\tau} u^{\mu} = D_{\tau} \Del u^{\mu} + \left[ \Del, D_{\tau} \right] u^{\mu} =
  \sg \lh D_{\tau}^2\, n^{\mu} - R_{\lb\nu\kg}^{\;\;\;\;\;\,\mu} u^{\lb} u^{\kg} n^{\nu} \rh = 0.
\label{3.3}
\ee
For example, we know all circular orbits in Schwarzschild space-time in analytical form. Then 
we can construct approximate non-circular (eccentric) solutions in the same plane by adding 
a geodesic deviation $\sg n^{\mu}$; generically this is an oscillating term as the test body 
moves periodically closer and farther from the central mass between periastron and apastron. 
As the period of this perturbation is in general different from the period of the circular orbit 
one starts from, the periastron and apastron will shift their azimuthal directions each turn. 
This phenomenon is well-known ever since it was calculated by Einstein for the planet 
Mercury in its orbit around the sun. It turns out that for Mercury the first-order deviation 
(\ref{3.2}) from motion on a circle with the period of Mercury's orbit actually already gives 
full numerical agreement with the observed periastron shift of 43 arcseconds per century 
\ct{kerner2001}. 

There is however no obstacle to include second and higher-order contributions in $\sg$ to 
the displaced geodesics $\xi^{\mu}(\sg,\tau)$. As $\Del$ creates a covariant displacement in 
the direction of $n$, but $n$ is not necessarily displaced parallel to itself (it is not required 
to be a tangent vector field to a family of geodesics crossing the world line $\xi^{\mu}(\tau)$), 
it follows that in general 
\be
\Del^2 \xi^{\mu} = \sg \Del n^{\mu} \equiv \sg^2 m^{\mu} \neq 0.
\label{3.4}
\ee
The geodesic family $\xi^{\mu}(\tau,\sg)$ is then parametrized to second order in $\sg$ by 
\be
\xi^{\mu}(\tau,\sg) = \xi_1^{\mu}(\tau) + \sg n^{\mu}(\tau) + 
 \frac{1}{2}\, \sg^2 \lh m^{\mu} - \Gam_{\lb\nu}^{\;\;\;\mu} n^{\lb} n^{\nu} \rh + ....
\label{3.5}
\ee
Using the properties of the vector field $n$ it is straightforward to generalize the derivation  
of eq.\ (\ref{3.3}) and show that $(n^{\mu}, m^{\mu}, ...)$ are solutions of a hierarchy 
of deviation equations~\ct{kerner2001}
\be
\ba{lll}
\dsp{ D_{\tau}^2 n^{\mu} - R_{\lb\nu\kg}^{\;\;\;\;\;\,\mu} u^{\kg} u^{\lb} n^{\nu} }& = & 0, \\
 & & \\
\dsp{ D_{\tau}^2 m^{\mu} - R_{\lb\nu\kg}^{\;\;\;\;\;\,\mu} u^{\kg} u^{\lb} m^{\nu} } & = & \dsp{
 \lh \nb_{\rg} R_{\kg\nu\lb}^{\;\;\;\;\;\,\mu} - \nb_{\kg} R_{\rg\lb\nu}^{\;\;\;\;\;\,\mu} \rh u^{\kg} u^{\lb} n^{\rg} n^{\nu} }\\
 \\
& & \dsp{ + 4 R_{\kg\nu\lb}^{\;\;\;\;\;\,\mu} u^{\kg} n^{\nu} D_{\tau} n^{\lb}, }\\
 \\
 ...
\ea
\label{3.6}
\ee
We note in passing, that whereas the first-order deviations $n^{\mu}$ provide information about the 
curvature of space-time through the Riemann tensor $R_{\kg\nu\lb\mu}$, the second-order
deviations provide further information about the gradient $\nb_{\rg} R_{\kg\nu\lb\mu}$ of 
the Riemann tensor. Thus with sufficient knowledge of families of geodesics in some domain of 
space-time one can reconstruct the Riemann tensor in the whole domain in terms of a Taylor series 
from the geodesic deviations w.r.t.\ a given geodesic \ct{szekeres1965,puetzfeld2015,philipp2016}. 

The procedure described here has been worked out for Schwarzschild space-time up to and 
including the second-order deviations \ct{kerner2001,koekoek2010,koekoek2011}. The results 
for orbits in the equatorial plane $\thg = \pi/2$ are summarized by the parametrized expansions
\be
\ba{lll}
\dsp{ t(\tau,\sg) = \xi_0^t(\sg) \tau + \sum_{n=1}^{\infty}\, \xi_n^t (\sg)\sin n \og(\sg) \tau, }\\
 \\
\dsp{ r(\tau,\sg) = \xi_0^r(\sg) + \sum_{n=1}^{\infty}\, \xi_n^r(\sg) \cos n \og(\sg) \tau, }\\
 \\
\dsp{ \vf(\tau,\sg) = \xi_0^{\vf}(\sg) \tau + \sum_{n = 1}^{\infty}\, \xi_n^{\vf}(\sg) \sin n \og(\sg) \tau. }\\
\ea
\label{3.7}
\ee
The coefficients $\xi_n^{\mu}(\sg)$ and the fundamental frequency $\og(\sg)$ are computed order by 
order in $\sg$ by solving the hierarchy of deviation equations (\ref{3.6}). Of course the terms of order 
$\sg^0$ represent the parameters of the circular parent orbit we encountered in eqs.\ (\ref{2.24}):
\be
\xi_0^t(0)= \frac{1}{\sqrt{1 - \frac{3GM}{R}}}, \hs{1} \xi_0^r(0) = R, \hs{1} 
\xi_0^{\vf}(0) = \sqrt{\frac{GM}{R^3}}\, \frac{1}{\sqrt{1 - \frac{3GM}{R}}}.
\label{3.8}
\ee
All other terms have non-trivial dependence on the expansion parameter $\sg$: 
\be
\ba{l}
\dsp{ \xi_0^{\mu}(\sg) = \xi_0^{\mu}(0) + \sg \rg_1^{\mu} + \frac{1}{2}\, \sg^2 \rg_2^{\mu} + ..., }\\
 \\
\dsp{ \xi_1^{\mu}(\sg) = \sg n_1^{\mu} + \frac{1}{2}\, \sg^2 n_2^{\mu} + ..., }\\
 \\
\dsp{ \xi_2^{\mu}(\sg) = \frac{1}{2}\, \sg^2 m_2^{\mu} + ...,}
\ea
\label{3.9}
\ee
each $\xi_n^{\mu}(\sg)$ contributing only terms of order $\sg^n$ and higher, whilst the angular 
frequency
\be
\og(\sg) = \og_0 + \sg \og_1 + ...
\label{3.10}
\ee
also depends on the order of approximation, reflecting the anharmonicity of the non-circular 
deviations. The explicit expressions for the coefficients $\rg_n^{\mu}$, $n_n^{\mu}$ and 
$m_n^{\mu}$ for $n = 1$ and $n = 2$ and the expressions for the frequencies $\og_0$, $\og_1$ 
are summarized in appendix B for the restricted case $\rg_n^r = 0$. This restriction implies, 
that we compare the non-circular orbits to a {\em fixed} circular orbit, although in general  
one might wish to adapt the circular reference orbit order by order in $\sg$ to improve 
convergence. The expressions for the unrestricted case can be found in ref.\ \ct{koekoek2010}.

Observe that the deviations are bound and periodic as long as the angular frequency $\og$ is 
real. For $R < 6GM$ equation (\ref{a.4}) in the appendix shows that it develops an imaginary part, 
indicating exponential run-away behaviour. Thus the circular orbit with $R = 6 GM$ is the innermost 
stable circular orbit (ISCO) for a simple test body in Schwarzschild space-time. 

Finally from the solutions of the geodesic deviations we can evaluate the constants of motion
$\ve$ and $\ell$ for the non-circular orbits by substitution of the solutions for $dt/d\tau$ and 
$d\vf/d\tau$ including first and second order deviations into the expressions (\ref{2.21}) and 
(\ref{2.22}). The resulting constants of motion are then also written as a series expansion 
in the deviation parameter $\sg$. Up to and including second order deviations the result is 
\be
\ve(\sg) = \ve_0 + \sg \ve_1 + \frac{1}{2}\, \sg^2 \ve_2 + ..., \hs{1}
\ell(\sg) = \ell_0 + \sg \ell_1 + \frac{1}{2}\, \sg^2 \ell_2 + ...,
\label{3.11}
\ee
where $\ve_0$ and $\ell_0$ are the energy and angular momentum per unit of mass for 
circular orbits: 
\be
\ve_0 = \frac{R - 2GM}{\sqrt{R(R - 3 GM)}}, \hs{1} \ell_0 = \sqrt{\frac{GMR^2}{R - 3GM}}.
\label{3.12}
\ee
In the restricted case $\rg^r_1 = 0$ the first-order corrections vanish: $\ve_1 = \ell_1 = 0$, 
but the second-order corrections do not: 
\be 
\ba{l} 
\dsp{ \ve_2 = - \frac{GM}{2 R^{7/2}}\, \frac{R^2 - 9 GMR + 6 (GM)^2}{(R - 3 GM)^{3/2}}, }\\
 \\
\dsp{ \ell_2 = - \frac{3\sqrt{GM}}{2R^2}\, \frac{(R - 2GM)(R - 7 GM)}{(R - 3 GM)^{3/2}}.  }
\ea
\label{3.13}
\ee
The most important aspect of these expressions is that all dependence on the proper time 
has disappeared and therefore they are true constants of motion indeed. This is a strong 
consistency check on the results listed in appendix B.

One of the useful aspects of the perturbative construction of orbits by the method of 
geodesic deviations is that they provide explicit and accurate expressions for the position 
of the test body as a function of the evolution parameter (proper time). Most results in the 
literature describe the time-dependence of the orbits only in an implicit way. The explicit 
time-dependent representation of the orbits of a test body in Schwarzschild space-time is 
especially useful in the computation of the emission of gravitational waves by extreme 
mass ratio (EMR) binaries: systems consisting of a compact stellar object, like a white dwarf, 
neutron star or stellar-mass black hole orbiting a supermassive black hole such as found 
in the center of many galaxies. Gravitational wave emission by EMR binaries has been 
studied in detail in ref.\ \ct{koekoek2011}.

\section{Spinning test bodies}

The motion of electrically neutral, non-spinning test bodies is represented by geodesics in
space-time. As soon as new degrees of freedom come into play the world line of a test body 
becomes non-geodesic. For example, an electrically charged particle in the presence of
electric or magnetic fields is subject to a Lorentz force in addition to the effect of curvature. 
Its world line will differ from that of a similar neutral particle which follows a geodesic. 

Similarly the effect of rotation represented by internal angular momentum (spin) will change 
the motion of test body in curved space-time even in the absence of other fields of force
\ct{mathisson1937}-\ct{tulczyjew1959}. One can think of this as a form of gravitational 
spin-orbit coupling. In this section we discuss the motion of spinning test bodies in more detail. 

There is an extended literature on spinning test particles; for a review see e.g.\ ref.\ 
\ct{steinhoff2011}. The most common approaches are based on a multipole expansion requiring 
a dynamical constraint as an implicit choice of the center of mass fixing the monopole term.  As this 
choice is not unique we have developed in refs.\ \ct{dambrosi2015a,vholten2015,dambrosi2015b} 
a different formalism which does not incorporate any constraints but only requires appropriate 
initial conditions to solve the equations of motion. In this formulation the spin degrees of freedom are described by a real 
anti-symmetric tensor field $\Sg^{\mu\nu}(\tau)$, which has six components defined on the world 
line. Three of these components can be represented equivalently by a space-like axial vector, a 
magnetic-type dipole moment which we identify with the spin proper. Similarly the other three 
components are described equivalently by a space-like real vector, an electric-type dipole 
moment which in the gravitational context is interpreted as a mass dipole. Denoting the 
four-velocity of the test body by the time-like unit vector $u^{\mu} = \dot{\xi}^{\mu}$, we can 
decompose the spin tensor $\Sg^{\mu\nu}$ accordingly as follows
\be
\Sg^{\mu\nu} =  - \frac{1}{\sqrt{-g}}\, \ve^{\mu\nu\kg\lb} u_{\kg} S_{\lb} + u^{\mu} Z^{\nu} - 
 u^{\nu} Z^{\mu},
\label{4.1}
\ee
with $S_{\mu}$ and $Z^{\mu}$ representing the proper spin and mass dipole components,
respectively. The decomposition (\ref{4.1}) can be inverted to get 
\be
S_{\mu} = \frac{1}{2}\, \sqrt{-g}\, \ve_{\mu\nu\kg\lb} u^{\nu} \Sg^{\kg\lb}, \hs{1}
Z^{\mu} = \Sg^{\mu\nu} u_{\nu}.
\label{4.2}
\ee
As mentioned both vectors are space-like by construction:  
\be
S \cdot u = 0, \hs{1} Z \cdot u = 0.
\label{4.3}
\ee
Hence each contains only three degrees of freedom, their time components vanishing in the 
rest frame $u = (1, 0, 0, 0)$. From these expressions it is infered directly that under the group 
of three-dimensional spatial rotations and parity transformations in the rest frame $Z$ is a real 
vector, whereas $S$ actually is an axial pseudo-vector. 

To obtain equations of motion for a spinning body we follow the same route as for a non-rotating 
body starting from an energy-momentum tensor and rederiving the results from an effective 
hamiltonian. Following \ct{vholten2015} the energy-momentum tensor is taken to have support 
on a world line $\xi^{\mu}(\tau)$  with
\be
\ba{lll}
T^{\mu\nu} & = & \dsp{ m \int d\tau\, u^{\mu} u^{\nu}\, \frac{1}{\sqrt{-g}}\, \del^4(x - \xi(\tau)) }\\
 & & \\
 & & \dsp{ + \frac{1}{2}\, \nb_{\lb} \int d\tau \lh u^{\mu} \Sg^{\nu\lb} + u^{\nu} \Sg^{\mu\lb} \rh 
 \frac{1}{\sqrt{-g}}\, \del^4(x - \xi(\tau)). }
\ea 
\label{4.4}
\ee
Its covariant divergence vanishes: $\nb_{\mu} T^{\mu\nu} = 0$, if 
\be
\ba{l} 
\dsp{ m D_{\tau} u^{\mu} = \dot{u}^{\mu} + \Gam_{\lb\nu}^{\;\;\;\mu} u^{\lb} u^{\nu} = 
 \frac{1}{2}\, \Sg^{\kg\lb} R_{\kg\lb\;\,\nu}^{\;\;\;\,\mu} u^{\nu}, }\\
 \\
\dsp{ D_{\tau} \Sg^{\mu\nu} = \dot{\Sg}^{\mu\nu} + u^{\lb} \Gam_{\lb\kg}^{\;\;\;\mu} \Sg^{\kg\nu} 
 + u^{\lb} \Gam_{\lb\kg}^{\;\;\;\nu} \Sg^{\mu\kg} = 0. }
\ea
\label{4.5}
\ee
Alternatively in the hamiltonian formulation we introduce a set of classical Poisson-Dirac brackets 
on the particle phase space defined by \ct{dambrosi2015a,dambrosi2015b}
\be
\ba{l}
\dsp{ \left\{ \xi^{\mu}, \pi_{\nu} \right\} = \del^{\mu}_{\nu}, \hs{2} 
\left\{ \pi_{\mu}, \pi_{\nu} \right\} = \frac{1}{2}\, \Sg^{\kg\lb} R_{\kg\lb\mu\nu}, }\\
 \\
\dsp{ \left\{ \Sg^{\mu\nu}, \pi_{\lb} \right\} = \Gam_{\lb\kg}^{\;\;\;\mu} \Sg^{\nu\kg} 
 - \Gam_{\lb\kg}^{\;\;\;\nu} \Sg^{\mu\kg}, }\\
 \\
\dsp{ \left\{ \Sg^{\mu\nu}, \Sg^{\kg\lb} \right\} = g^{\mu\kg} \Sg^{\nu\lb} - g^{\mu\lb} \Sg^{\nu\kg} 
 - g^{\nu\kg} \Sg^{\mu\lb} + g^{\nu\lb} \Sg^{\mu\kg}. }
\ea
\label{4.6}
\ee
These brackets have several important features. First, they define an algebra with structure functions 
rather than structure constants. Moreover these structure functions encode all of the usual space-time
geometry: the metric, the connection and the Riemann curvature tensor. In addition the properties of
these geometric objects guarantee that the Jacobi identities for cyclic triple brackets are satisfied by 
this Poisson-Dirac algebra without specifying the dynamics of the system. Therefore the bracket algebra
is consistent independently of the choice of hamiltonian. In particular the free hamiltonian (\ref{2.9}):
\[
H = \frac{1}{2m}\, g^{\mu\nu}[\xi]\, \pi_{\mu} \pi_{\nu},
\]
generates the same equations of motion (\ref{4.5}) as derived from the energy-momentum tensor, 
providing in addition the identification of $\pi_{\mu}$ with the kinetic momentum (\ref{2.10.1}):
\[
\pi_{\mu} = m g_{\mu\nu} u^{\nu}.
\]
To solve the equations of motion (\ref{4.5}) it is again convenient to first identify constants of 
motion. In this case there are 3 universal constants of motion, independent of the specific 
geometry: the hamiltonian itself:
\[ 
H = - \frac{m}{2},
\]
as long as it is proper-time independent; and two spin invariants:
\be
\ba{l}
\dsp{ I = \frac{1}{2}\, g_{\mu\kg} g_{\nu\lb} \Sg^{\mu\nu} \Sg^{\kg\lb} = S^2 - Z^2, }\\
 \\
\dsp{ D = \frac{1}{8}\, \sqrt{-g}\, \ve_{\mu\nu\kg\lb} \Sg^{\mu\nu} \Sg^{\kg\lb} = S \cdot Z. }
\ea
\label{4.7}
\ee
Clearly $I$ is a real scalar whilst $D$ is a pseudoscalar under three-dimensional spatial 
rotations and parity transformations. 

In addition there may exist other constants of motion connected with symmetries of the 
background space time. The construction in equations (\ref{2.12}), (\ref{2.13}) based on
the presence of Killing vectors can be generalized immediately to include spin 
\ct{ehlers1977,ruediger1981}, as follows: a quantity
\be
J[\xi,\pi,\Sg] = \ag^{\mu}[\xi]\, \pi_{\mu} + \frac{1}{2}\, \bg_{\mu\nu}[\xi]\, \Sg^{\mu\nu} 
\label{4.8}
\ee
is a constant of motion if $\ag^{\mu}$ is a Killing vector and $\bg_{\mu\nu}$ its gradient: 
\be
\nb_{\mu} \ag_{\nu} + \nb_{\nu} \ag_{\mu} = 0, \hs{2}
\bg_{\mu\nu} = - \bg_{\nu\mu} = \nb_{\mu} \ag_{\nu}.
\label{4.9}
\ee
Observe that the symmetric part of $\bg_{\mu\nu}$ vanishes by construction as a result 
of the Killing condition; this property also implies the identity  
\be
\nb_{\lb} \bg_{\mu\nu} = R_{\mu\nu\lb}^{\;\;\;\;\;\,\kg}\, \ag_{\kg}, 
\label{4.1)}
\ee
which is necessary to show that $J$ is a constant of motion. 

\section{Spinning test bodies in Schwarzschild space-time} 

Applying the general formalism above to the case of Schwarzschild space-time we first 
construct the generalization of the constants of motion (\ref{2.14}) and (\ref{2.16}), to wit
the kinetic energy based on the Killing vector of time-translation invariance:
\be
E = - \pi_t  - \frac{GM}{r^2}\, \Sg^{tr} = m \lh 1 - \frac{2GM}{r} \rh u^t - \frac{GM}{r^2}\, \Sg^{tr} , 
\label{5.1}
\ee
and the total angular momentum based on the Killing vectors of invariance under rotations:
\be
\ba{lll}
J_1 & = & \dsp{ - \sin \vf\, \pi_{\thg} - \ctg \thg \cos \vf\, \pi_{\vf} }\\
 & & \\
 & & \dsp{ - r \sin \vf\, \Sg^{r\thg} - r \sin \thg \cos \thg \cos \vf\, \Sg^{r\vf} + r^2 \sin^2 \thg \cos \vf\, \Sg^{\thg\vf}, }\\
 & & \\
J_2 & = & \dsp{ \cos \vf\, \pi_{\thg} - \ctg \thg \sin \vf\, \pi_{\vf} }\\
 & & \\
 & & \dsp{ + r \cos \vf\, \Sg^{r\thg} - r \sin \thg \cos \thg \sin \vf\, \Sg^{r\vf} + r^2 \sin^2 \thg \sin \vf\, \Sg^{\thg\vf}, }\\
 & & \\
J_3 & = & \dsp{ \pi_{\vf} + r \sin^2 \thg\, \Sg^{r\vf} + r^2 \sin \thg \cos \thg\, \Sg^{\thg\vf}. }
\ea
\label{5.2}
\ee
As for the spinless test bodies we can orient the co-ordinate system such that the $z$-axis coincides with 
the direction of the total angular momentum:
\be
\bfJ = (0. 0, J), \hs{2} J = mr^2 u^{\vf} + r \Sg^{r\vf}.
\label{5.3}
\ee
The total spin $J$ is now composed of the contributions from orbital angular momentum and from spin 
in the $z$-direction. The contributions in the perpendicular directions must then cancel. This is expressed 
by the conditions
\be
\Sg^{r\thg} = - m r u^{\thg}, \hs{2} \Sg^{\thg\vf} = \frac{J}{r^2}\, \ctg \thg.
\label{5.4}
\ee
As a transverse component of spin must be compensated by a transverse component of orbital 
angular momentum, the spin proper can precess only if the orbital angular momentum precesses 
as well. Orbits can therefore be strictly planar if and only if the spin and orbital angular momentum 
are permanently aligned. 

With that restriction it is still possible to find planar and even circular orbits. They necessarily 
require all $\thg$-components of the spin tensor to vanish: 
\be
\Sg^{t\thg} = \Sg^{r\thg} = \Sg^{\thg\vf} = 0,
\label{5.5}
\ee
and therefore $D = S \cdot Z = 0$ identically. Moreover for circular orbits $r = R$ is a constant,
and $u^r = \dot{u}^r = 0$. This is sufficient to fix the motion of a spinning test body, in the sense 
that we can derive equations for the time dilation $u^t$ and the angular velocity $u^{\vf}$ in 
terms of the energy per unit of mass $\ve = E/m$ and the total angular momentum per unit 
of mass $\eta = J/m$:
\be
\ba{l}
\ve R^2 \lh 1 - u^{t\,2} \rh = R \lh R- 3 GM \rh u^t - \lh R^2 - 3 GM R + 3 (GM)^2 \rh u^{t\,3}, \\
 \\
\dsp{ \eta \lh 2 GM + R^3 u^{\vf\,2} \rh = R^3 u^{\vf} \left[ 1 - \frac{R^3 u^{\vf\,2}}{GM} 
 \lh 1 - \frac{6GM}{R} + \frac{6(GM)^2}{R^2} \rh \right]. }
\ea
\label{5.6}
\ee
Finally the solutions also have to satisfy the hamiltonian constraint 
\be
\lh 1 - \frac{2GM}{R} \rh u^{t\,2} = 1 + R^2 u^{\vf\,2} \geq 1.
\label{5.7}
\ee
Therefore of the five parameters $(\ve, \eta, R,u^t, u^{\vf})$ characterizing a circular obit only 
two are independent. Implicitly by eqs.\ (\ref{5.1}) and (\ref{5.3}) they also determine the 
non-vanishing components of the spin tensor; these are related to the velocity components by
\be
\ba{l}
\dsp{ \frac{1}{m}\, \Sg^{tr} \lh u^{t\,2} - 1 \rh = u^t \left[ (R - 3GM) u^{t\,2} - R \right], }\\
 \\
\dsp{ \frac{1}{m}\, \Sg^{r\vf} \lh 2GM + R^3 u^{\vf\,2} \rh = 
   (R - 2GM) Ru^{\vf} \left[ 1 - \lh \frac{R}{GM} - 3 \rh R^2 u^{\vf\,2} \right]. }
\ea
\label{5.8}
\ee
Having a complete parametrization of circular orbits for spinning test bodies it is straightforward 
to construct non-circular orbits by the method of world-line deviations, a direct generalization of the 
geodesic-deviation procedure explained in sect.\ 3. First the deviations $(\del \xi, \del u, \del \Sg)$ 
near a given reference world line can be recombined in covariant expressions
\be
\ba{l} 
\Del \xi^{\mu} = \del \xi^{\mu}, \hs{1} 
\Del u^{\mu} = \del u^{\mu} + \del \xi^{\lb} \Gam_{\lb\nu}^{\;\;\;\mu} u^{\nu}, \\ 
 \\
\Del \Sg^{\mu\nu} = \del \Sg^{\mu\nu} + \del \xi^{\lb} \Gam_{\lb\kg}^{\;\;\;\mu} \Sg^{\kg\nu} 
 + \del \xi^{\lb} \Gam_{\lb\kg}^{\;\;\;\nu} \Sg^{\mu\kg}.
\ea
\label{5.9}
\ee
Note that there is no {\em a priori} relation between the variations $\Del \xi^{\mu}$ and 
$\Del \Sg^{\mu\nu}$, but all variations are linked by the first-order world-line deviation equations 
\be
\ba{l}
D_{\tau} \Del \xi^{\mu} = \Del u^{\mu}, \\
 \\
\dsp{ D_{\tau} \Del u^{\mu} - R_{\nu\kg\;\;\lb}^{\;\;\;\,\mu} u^{\kg} u^{\lb} \Del \xi^{\nu} = 
 \frac{1}{2m}\, \Sg^{\rg\sg} R_{\rg\sg\;\;\nu}^{\;\;\;\,\mu} \Del u^{\nu} + 
 \frac{1}{2m}\, \Del \Sg^{\rg\sg} R_{\rg\sg\;\;\nu}^{\;\;\;\,\mu} u^{\nu} }\\
 \\
\dsp{ \hs{12} +\, \frac{1}{2m}\, \Sg^{\rg\sg} \nb_{\lb} R_{\rg\sg\;\;\nu}^{\;\;\;\,\mu} u^{\nu} \Del \xi^{\lb}, }\\
 \\
\dsp{ D_{\tau} \Del \Sg^{\mu\nu} = \lh R_{\kg\lb\sg}^{\;\;\;\;\;\,\mu} \Sg^{\sg\nu} + 
 R_{\kg\lb\sg}^{\;\;\;\;\;\,\nu} \Sg^{\mu\sg} \rh u^{\kg} \Del \xi^{\lb}.  }
\ea
\label{5.10}
\ee
Here we will consider in particular planar non-circular orbits. The special conditions (\ref{5.5}) 
then still apply. The only new degrees of freedom are the radial velocity $u^r$ and the mass 
dipole moment $\Sg^{t\vf}$ which become non-zero. Above we observed that for circular 
orbits the parameters $\ve$ and $\eta$, representing energy and total angular momentum 
per unit of mass of the test body, can be varied independently even for circular orbits, e.g.\ 
by adjusting $z$-component of the spin. Therefore there is always a circular orbit with 
the same $\ve$ and $\eta$ as the non-circular orbit we wish to construct. For simplicity we 
will take this circular orbit as the reference orbit. Then the covariant deviation equations 
(\ref{5.10}) for near-circular orbits in Schwarzschild space-time reduce to a set of 
{\em homogeneous} linear differential equations for the deviations as functions of proper time 
$\tau$. Now as the spin-tensor deviations are independent of the orbital deviations there 
are actually two independent types of solutions of the coupled deviation equations. Indeed, 
the first-order approximation for the world lines of spinning bodies takes the form
\be
\ba{l}
t(\tau) = u^t \tau + \sg_+ n^t_+ \sin \og_+ (\tau - \tau_+) + \sg_- n^t_- \sin \og_- (\tau - \tau_-), \\
 \\
\vf(\tau) = u^{\vf} \tau + \sg_+ n^{\vf}_+  \sin \og_+ (\tau - \tau_+) + \sg_- n^{\vf}_- \sin \og_- (\tau - \tau_-), \\ 
 \\
r(\tau) = R + \sg_+ n^r_+ \cos \og_+ (\tau - \tau_+) + \sg_- n^r_- \cos \og_- (\tau - \tau_-),\\
 \\ 
\Sg^{tr}(\tau) = \Sg^{tr}_0 + \sg_+ N^{tr}_+ \cos \og_+ (\tau - \tau_+) 
 + \sg_- N^{tr}_- \cos \og_-(\tau - \tau_-), \\
 \\
\Sg^{r\vf}(\tau) = \Sg^{r\vf}_0 + \sg_+ N^{r\vf}_+ \cos \og_+ (\tau - \tau_+) 
 + \sg_- N^{r\vf}_- \cos \og_-(\tau - \tau_-), \\
 \\
\Sg^{t\vf}(\tau) = \sg_+ N^{t\vf}_+ \sin \og_+ (\tau - \tau_+) + \sg_- N^{t\vf}_- \sin \og_- (\tau - \tau_-),
\ea
\label{5.11}
\ee
where $\sg_{\pm}$ are the two independent expansion parameters for the different deviation modes, 
which are characterized by fundamental angular frequencies $\og_{\pm}$. The corresponding 
amplitudes are denoted by $(n^{\mu}_{\pm}, N^{\mu\nu}_{\pm})$ for the orbital and spin-tensor 
deviations respectively. The explicit expressions for the frequencies and amplitudes are collected 
in appendix C.

Finally the lowest order terms $\Sg^{tr}_0$ and $\Sg^{r\vf}_0$ for the spin tensor components 
are the circular-orbit values satisfying equations (\ref{5.8}), and $\tau_{\pm}$ are constants
of integration, one for each type of deviation, which fix their initial values. For a complete derivation 
of these results I refer to ref.\ \ct{dambrosi2015b}.

\section{Stability of orbits and the ISCO}

In section 3 it was shown that for radial co-ordinates $R < 6GM$ the deviations from circular 
orbits show exponential runaway behaviour. Therefore $R = 6 GM$ is the innermost stable 
circular orbit. In the case of spinning particles we can similarly investigate the solutions (\ref{5.11}) 
of the equations for deviations from circular orbits for instabilities and determine the existence 
of an ISCO for different values of the $z$-component of spin per unit of mass
\[
\sg = \frac{R \Sg^{r\vf}}{m},
\]
which is the spin-contribution to $\eta$.

\begin{figure}
\begin{center}
\scalebox{1}{\includegraphics{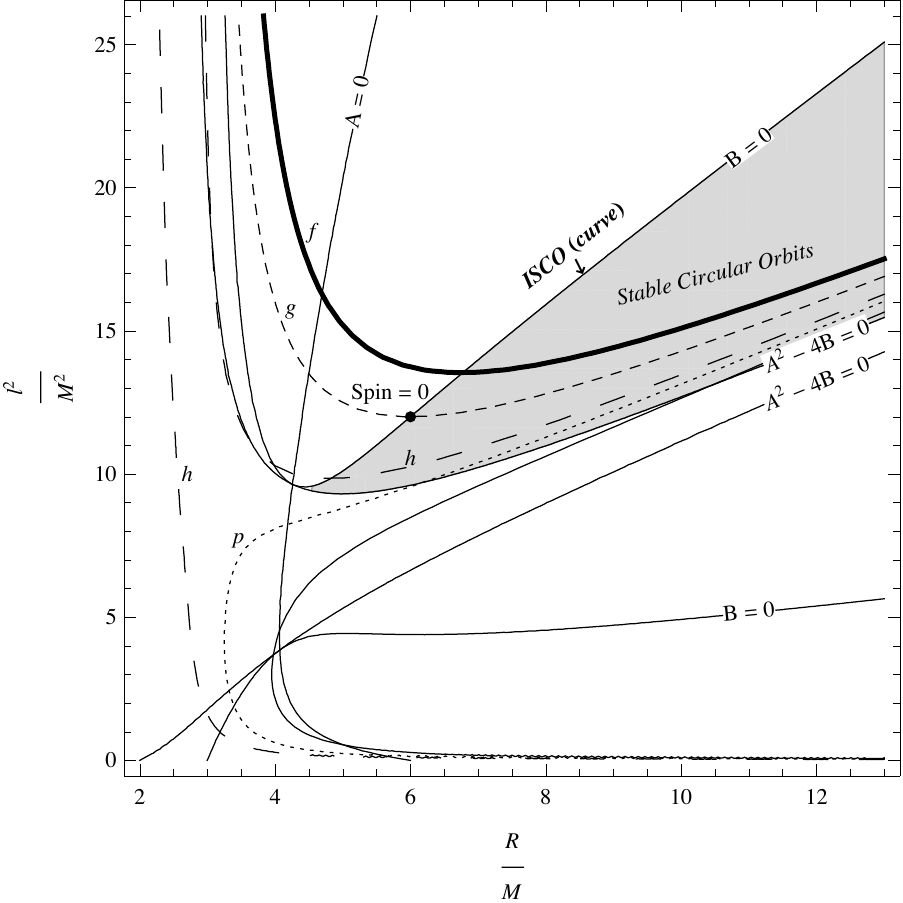}}
\end{center}
\caption{Region of stability of circular orbits in the $R$-$\ell^2$ plane.}
\label{fig:1}
\end{figure}

The stable deviations (\ref{5.11}) are characterized by real-valued angular frequencies $\og_{\pm}$.
If any one of these frequencies develops an imaginary part runaway behaviour sets in and 
the circular orbits become unstable. The frequencies themselves are given by the expressions
(\ref{ab.3}) in appendix C:
\[
\og^2_{\pm} = \frac{1}{2}\, \lh A \pm \sqrt{A^2 - 4B} \rh,
\]
where $A$ and $B$ represent long expressions in terms of the parameters of the circular 
reference orbit. For $\og_{\pm}$ to be real the square root on the right-hand side must be 
real, and the whole expression must be nonnegative as it represents a real square. 
This results in the following inequalities 
\be
A \geq 0 \hs{1} \mbox{and} \hs{1} 0 \leq 4B \leq A^2.
\label{6.1}
\ee
These conditions are plotted in figure \ref{fig:1} as a function of the radial co-ordinate $R$ and the 
orbital angular momentum
\[
\ell = R^2 u^{\vf} = \eta - \sg,
\]

\begin{figure}
\begin{center}
\scalebox{0.9}{\includegraphics{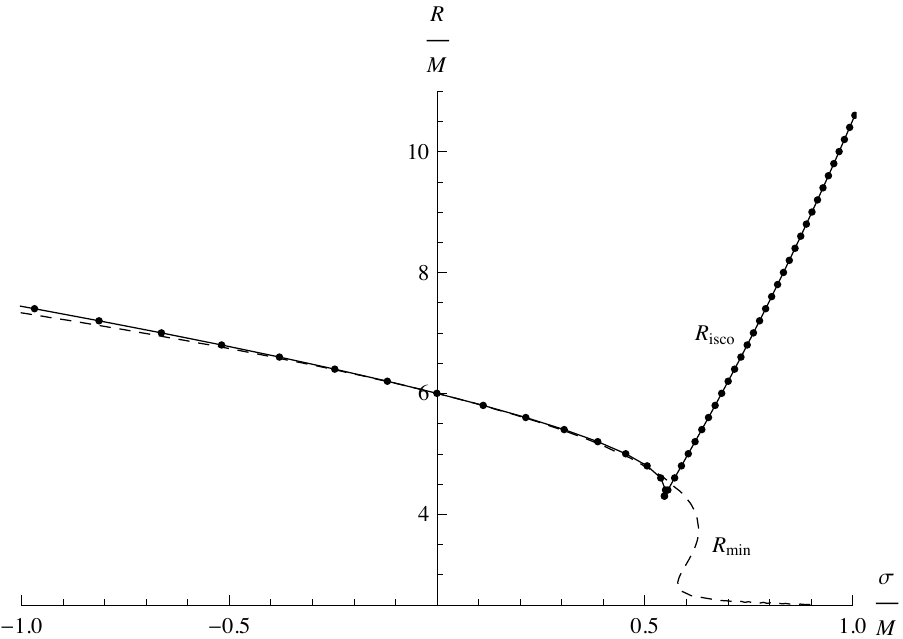}}
\end{center}
\caption{ ISCO radius $R$ as function the spin parameter $\sg$. 
 The dashed line represents the values of $R$ for which 
 the orbital angular momentum reaches its minimum.}
\label{fig:2}
\end{figure}

\nit
both measured in units of $GM$. The curves labeled {\em f, g, h} represent lines of constant 
$\sg$. In particular {\em g} is the curve for spinless test bodies ($\sg = 0$); it leaves the region 
of stability at $R = 6 GM$, as expected. For retrograde spin values $\sg < 0$ as represented 
by the curve {\em f} the ISCO is reached earlier, whilst prograde spin values $\sg > 0$ as 
represented by the curve {\em h} stabilize circular orbits in a range of values $R < 6 GM$. 
From these results one can infer the radial co-ordinate $R$ of the ISCO as a function of the 
spin parameter $\sg$, as plotted in figure \ref{fig:2}.

The steep line for spin values $\sg > 0.55$ has been included for completeness; here 
the upper limit on $B$ in inequality (\ref{6.1}) takes over from the condition $B > 0$
as the main stability criterion. However these large spin values are physically unrealistic
as they can only be obtained in cases where the test-body limit is not applicable, such as 
binary black holes of comparable mass. Also plotted in figure 6.2 is the curve obtained 
by minimizing the orbital angular momentum $\ell$ as a function of $R$ at fixed spin. 
Cleary the two curves largely coincide. 

\section{Non-minimal hamiltonian dynamics of spinning test bodies} 

The motion of test bodies has been modeled so far using the minimal hamiltonian (\ref{2.9}).
However, it is not difficult to construct more complicated hamiltonians to model test bodies with 
additional interactions such as spin-curvature couplings. As the Dirac-Poisson brackets (\ref{4.6}) 
are closed and model independent the equations of motion can be derived in straightforward 
fashion for any such extended hamiltonian. For example, one can include Stern-Gerlach type of 
interactions as discussed in refs.\ \ct{khriplovich1999,dambrosi2015a,dambrosi2015b}. In this 
case the extended test-body hamiltonian is
\be
H =  \frac{1}{2m}\, g^{\mu\nu}[\xi]\, \pi_{\mu} \pi_{\nu} + 
  \frac{\kg}{4}\, R_{\mu\nu\kg\lb}[\xi]\, \Sg^{\mu\nu} \Sg^{\kg\lb}.
\label{7.1}
\ee
In terms of the four-velocity $u^{\mu} = \dot{\xi}^{\mu}$ the corresponding equations of motion read
\be
\ba{l}
\pi_{\mu} = m g_{\mu\nu} u^{\nu}, \\
 \\
\dsp{ m D_{\tau} u^{\mu} = \frac{1}{2}\, \Sg^{\kg\lb} R_{\kg\lb\;\;\nu}^{\;\;\;\,\mu} u^{\nu} 
 - \frac{\kg}{4}\, \Sg^{\kg\lb} \Sg^{\rg\sg} \nb_{\mu} R_{\kg\lb\rg\sg}, }\\
 \\
\dsp{ D_{\tau} \Sg^{\mu\nu} = - \kg \Sg^{\rg\sg} \lh R_{\rg\sg\;\;\lb}^{\;\;\;\,\mu} \Sg^{\lb\nu} 
 + R_{\rg\sg\;\;\lb}^{\;\;\;\,\nu} \Sg^{\mu\lb} \rh. } 
\ea
\label{7.2}
\ee
As in the minimal case these equations can also be derived by requiring the vanishing of the covariant
divergence of a suitable energy-momentum tensor \ct{vholten2015}
\be
\ba{lll}
T^{\mu\nu} & = & \dsp{ T^{\mu\nu}_{min} + \frac{\kg}{4}\, \nb_{\kg} \nb_{\lb}\, \int d\tau
 \lh \Sg^{\mu\lb} \Sg^{\kg\nu} + \Sg^{\nu\lb} \Sg^{\mu\kg} \rh  \frac{1}{\sqrt{-g}}\, \del^4(x - \xi(\tau)) }\\
 & & \\
 & & \dsp{ +\, \frac{\kg}{4}\, \int d\tau\, \Sg^{\rg\sg} \lh R_{\rg\sg\lb}^{\;\;\;\;\;\,\mu} \Sg^{\lb\nu} 
  + R_{\rg\sg\lb}^{\;\;\;\;\;\,\nu} \Sg^{\lb\mu} \rh \frac{1}{\sqrt{-g}}\, \del^4(x - \xi(\tau)). }
\ea
\label{7.3}
\ee
Here $T^{\mu\nu}_{min}$ is the energy-momentum tensor (\ref{4.4}) of a spinning test body with minimal 
dynamics. 

Remarkably all conservation laws for spinning bodies we derived in the minimal case carry over to 
the case with Stern-Gerlach interactions. In particular any constant of motion (\ref{4.8}), (\ref{4.9}) 
associated with a Killing vector $\ag^{\mu}$ is also conserved by the Stern-Gerlach terms in the 
hamiltonian: 
\be
\frac{\kg}{4}\, \left\{ J, R_{\mu\nu\kg\lb} \Sg^{\mu\nu} \Sg^{\kg\lb} \right\} = 0.
\label{7.4}
\ee
For example, in a static and spherically symmetric background like Schwarz\-schild or 
Reissner-Nordstr{\o}m space-time the kinetic energy $E$ and the angular momentum 3-vector 
$\bfJ$ given by equations (\ref{5.1}) and (\ref{5.2}) are again conserved. 

This form of non-minimal hamiltonian dynamics predicts some interesting effects. In particular, 
as the hamiltonian is a constant of motion which by evaluation in a curvature-free region is seen 
to be expressed in terms of the inertial mass by $H = - m/2$, the hamiltonian constraint gets 
modified to read
\be
g^{\mu\nu} \pi_{\mu} \pi^{\nu} + \frac{\kg m}{2}\, R_{\mu\nu\kg\lb}\, \Sg^{\mu\nu} \Sg^{\kg\lb} 
  + m^2 = 0.
\label{7.5}
\ee
Then the four-velocity is no longer normalized to be a time-like unit vector; instead the time-like 
unit vector tangent to the world line actually is
\be
n^{\mu} = \frac{u^{\mu}}{\lh 1 + \frac{\kg}{2m}\, R_{\mu\nu\kg\lb}\, \Sg^{\mu\nu} \Sg^{\kg\lb} \rh^{1/2}}.
\label{7.6}
\ee
Considering a particle at rest:
\be
g_{tt} \lh \frac{dt}{d\tau} \rh^2 = 1 + \frac{\kg}{2m}\, R_{\mu\nu\kg\lb}\, \Sg^{\mu\nu} \Sg^{\kg\lb},
\label{7.7}
\ee
this is seen to imply that the spin-curvature coupling represents an additional source of 
gravitational time-dilation. A similar effect related to spinning particles interacting with 
electromagnetic fields was conjectured in refs.\ \ct{vholten1990,vholten1992}. 

\section{Final remarks} 

The motion of test bodies carrying a finite number of relevant degrees of freedom like momentum, 
spin or charge can be represented by world lines in space-time to the extent that we can assign 
them a well-defined position and that their back reaction on space-time geometry can be neglected. 
Convenient position co-ordinates are not necessarily those of a center of mass (or for that matter 
a center of charge) in the local rest frame, as the example of spinning test bodies shows. In that 
case we find it preferable to associate the world line of free particles with the line on which the spin 
tensor is covariantly constant. The mass dipole moment can then be taken to represent the effective 
position of the mass with respect to that world line. 

This is also clear from the corresponding energy-momentum tensor which receives contributions 
from both the spin proper and the mass dipole. In a next step this can be used to compute the back 
reaction of the test body on the space-time geometry as discussed in the simple example in section 1. 
In general this procedure also includes determining the self-force and the gravitational waves emitted 
by test bodies in the specific background under discussion \ct{koekoek2011,dambrosi2014}. 

As another application it has been shown in the literature how the motion of test bodies can be used 
to reconstruct the geometry of space-time \ct{puetzfeld2015}. Simple geodesic motion of a sufficient 
number of test bodies allows one to determine the curvature at a point in space by measuring the 
geodesic deviations in its neighborhood. By including higher-order corrections as in eqs.\ (\ref{3.5}), 
(\ref{3.6}) one could also determine the derivatives of the curvature to obtain the curvature in a region 
around the point of interest. As equations (\ref{5.10}) show, an alternative method is to measure 
first-order world-line deviations of spinning test bodies, which also depend on the gradient of the 
Riemann curvature tensor. 
\vs{4} 

\nit
{\bf Acknowledgment} \\
I am indebted to Richard Kerner, Roberto Collistete jr., Gideon Koekoek, Giuseppe d'Ambrosi, 
S.\ Satish Kumar and Jorinde van de Vis for pleasant and informative discussions and collaboration 
on various aspects of the topics discussed. This work is supported by the Foundation for Fundamental 
Research of Matter (FOM) in the Netherlands. 
\vs{1}
 
\appendix

\section{Observer-dependence of the center of mass in relativity} 

\begin{figure}[h]
\begin{center}
\scalebox{0.28}{\includegraphics{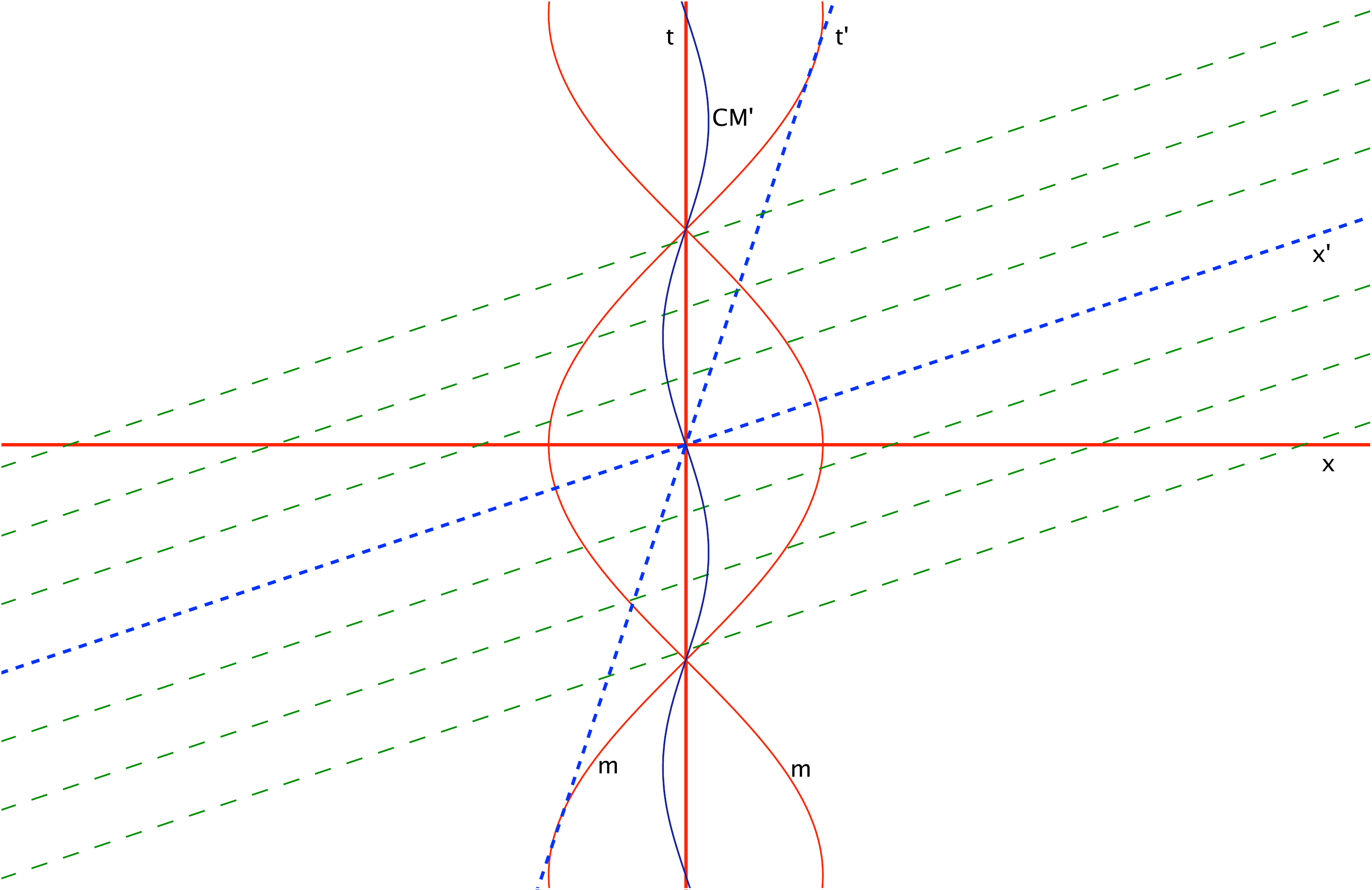}}
\end{center}
\caption{ The world line of the center of mass of two equal masses (e.g., a binary star system)
               in circular orbit with respect to a stationary observer on the axis of orbital angular 
               momentum is represented by the line $x = 0$. The world line of the center of 
               mass with respect to an observer in an inertial frame $(t',x')$ moving at constant  
               velocity along the $x$-axis is represented by the oscillating curve labeled CM$'$. }
\label{fig:3}
\end{figure}

To illustrate the observer-dependence of the center of mass of an extended body we 
consider a simple example: the motion of two equal test bodies revolving at constant 
angular velocity in Minkowski space on a circular orbit around an observer located 
in the origin of an inertial frame with cartesian co-ordinates $(t,x,y,z)$. The plane of 
the orbit is taken to be the $x$-$y$-plane. In fig.\ \ref{fig:3} we plot the projection 
of the world lines in the $x$-$t$-plane, represented by the two widely oscillating curves.
At any moment the two test bodies are at equal distance to the observer and in opposite 
phase with respect to the origin. In this frame the center of mass is located at the 
origin $x = 0$ and moves in a straight line along the $t$-axis in space-time. 

A second observer in another inertial frame $(t',x',y',z')$ moving with constant velocity $v$ 
in the positive $x$-direction has a different notion of simultaneity, as defined by the 
appropriate Lorentz transformation. The lines $t' =$ constant are represented by the 
dashed slant lines parallel to the $x'$-axis. In the limit of large masses and slow rotation 
the center of mass CM$'$ with respect to this moving frame is located halfway between 
the masses at fixed
\vs{.5}

\nit
time $t'$. The world line of CM$'$ is now represented by the single curve oscillating at smaller 
amplitude around the line $x = 0$ in the original frame.  In fact for the observer in relative motion 
CM$'$ moves in the negative $x'$-direction while oscillating around the line $x' = - vt'$.

It is obvious that in curved space-time the notion of simultaneity is further complicated 
because of the non-existence of global inertial frames, resulting in additional distortions 
of the world line CM$'$ with respect to the world line in the local inertial frame $(t,x,y,z)$ 
fixed to the center of rotation.

\section{Coefficients for geodesic deviations in \\ Schwarzschild geometry} 

The coefficients for the deviations of bound equatorial orbits w.r.t.\ parent circular orbits have 
been calculated for Schwarzschild space-time up to second order; with the restriction 
$\rg^r_1 = \rg^r_2 = 0$ explained in the main text one gets the following results 
\ct{kerner2001}: \\
a.\ Secular terms: 
\be
\ba{ll}
\rg_1^t = 0, & \dsp{ \rg_2^t = \frac{3GM}{2R^{5/2}} \frac{R + GM}{\lh R - 3GM \rh^{3/2}}, }\\
 & \\
\rg_1^{\vf} = 0, & \dsp{ \rg_2^{\vf} = \frac{3}{2R^{7/2}} \sqrt{\frac{GM}{R}} 
 \frac{\lh R - 2GM \rh \lh R + GM \rh}{\lh R - 3GM \rh^{3/2}}. }
\ea
\label{a.1}
\ee
b.\ First-order periodic terms:
\be
\ba{ll}
\dsp{ n_1^t = - \sqrt{\frac{4GM R}{ \lh R - 2GM \rh \lh R - 6GM \rh}}, }& n_2^t = 0, \\ 
 & \\
\dsp{ n_1^r = \sqrt{ 1 - \frac{2GM}{R}}, }& n_2^r = 0, \\
 & \\
\dsp{ n_1^{\vf} = - \frac{2}{R} \sqrt{\frac{R - 2 GM}{R - 6GM}}, }& n_2^{\vf} = 0.
\ea
\label{a.2}
\ee
c.\ Second order periodic terms: 
\be
\ba{l}
\dsp{ m_2^t =  \sqrt{\frac{GM}{R^2}}\, 
 \frac{2R^2 - 15 GMR + 14 (GM)^2}{\lh R - 2GM \rh \lh R - 6 GM \rh^{3/2}}, }\\
 \\
\dsp{ m_2^r = - \frac{1}{R^2}\, \frac{\lh R - 2GM \rh \lh R - 7 GM \rh}{R - 6 GM}, }\\
 \\
\dsp{ m_2^{\vf} = \frac{1}{2 R^{5/2}}\, 
 \frac{\lh R - 2 GM \rh \lh 5R - 32 GM \rh}{\lh R - 6GM \rh^{3/2}}. }
\ea
\label{a.3}
\ee
d.\ Angular frequency: 
\be
\og_0 = \sqrt{ \frac{GM}{R^3} \frac{R - 6 GM}{R - 3 GM}}, \hs{1} \og_1 = 0.
\label{a.4}
\ee
In the non-restricted case with $\rg_1^r \neq 0$ also the coefficients $\rg_1^t$, 
$\rg_1^{\vf}$, $n_2^{\mu}$ and $\og_1$ all become non-zero as well \ct{koekoek2010}.

\section{Coefficients for spinning world-line deviations in Schwarzschild geometry} 

The first-order planar deviations of circular orbits of spinning particles for constant energy 
and total angular momentum in Schwarzschild space-time are expressed conveniently in 
terms of the following combinations of orbital and spin parameters \ct{dambrosi2015b} 
\be
\ba{ll}
\alpha = \dsp{ \frac{2(R-GM)}{R(R-2GM)}\,  u^t - \frac{2\ve}{R-2GM}, }& 
\beta = \dsp{ -\, \frac{GM u^\vf}{mR}, }\\
 & \\
\gamma = \dsp{ \frac{2R-5GM}{R(R - 2GM)}\, u^\vf + \frac{GM \eta}{R^3(R-2GM)}, }&
\zeta = \dsp{ -\, \frac{GM(R-2GM)}{mR^4}\, u^t, }\\
 & \\
\kappa = \dsp{ -  \frac{2(R-2GM)}{R^2}\, \ve, }& \lambda = \dsp{ 2R u^{\vf} - \frac{GM \eta}{R^2}, }\\
\ea
\label{ab.1}
\ee
and  
\be
\ba{lll}
\mu & = & \dsp{ -\frac{2(R-3GM)}{R^3} +\, \frac{2(R-4GM)}{R^3}\, \ve u^t 
 +\, u^{\vf\,2} + \frac{2GM}{R^3}\, \eta u^{\vf}, }\\
 & & \\
\nu & = & \dsp{ \frac{(R-GM)(R - 3 GM)}{R-2GM}\, m u^{\vf} + 
  \frac{GM m\eta}{R^2(R-2GM)}, }\\ 
 & & \\
\sigma & = & \dsp{ \frac{(R- GM)(R-3GM)}{GM(R-2GM)}\, m u^t 
 -\, \frac{mR\ve}{GM}, }\\
 & & \\
\chi & = & \dsp{ \frac{(R^2-4GMR+5(GM)^2)}{GM(R-2GM)^2}\, m u^{\vf} u^t 
  -\, \frac{GM(3R-4GM)}{R^3(R-2GM)^2}\, m \eta u^t -\, \frac{m \ve u^{\vf}}{GM}. }
\ea
\label{ab.2}
\ee
With these definitions the frequencies of the first-order planar deviations are 
\be 
\ba{l}
\dsp{ \og^2_{\pm} = \frac{1}{2} \lh A \pm \sqrt{A^2 - 4B} \rh, }\\
 \\
A = \mu - \ag \kg - \bg \nu - \gam \lb - \zg \sg, \\
  \\
B = \bg \lh \kg \chi - \mu\nu + \gam (\lb \nu - \kg \sg) \rh + 
  \zg \lh \lb \chi - \mu \sg - \ag (\lb \nu - \kg \sg) \rh,
\ea
\label{ab.3}
\ee 
whilst the amplitudes are given by
\be
\ba{l}
n^t_{\pm} = \lb (\bg \gam - \ag \zg) + \bg (\og^2_{\pm} - \mu), \\
 \\
n^{\vf}_{\pm} = - \kg (\bg \gam - \ag \zg) + \zg (\og^2_{\pm} - \mu), \\
 \\
n^r_{\pm} = \og_{\pm} (\bg \kg + \zg \lb), 
\ea
\label{ab.4}
\ee
and
\be
\ba{l}
\dsp{ N^{tr}_{\pm} = \frac{m \og_{\pm} R^2}{GM} \lh 1 - \frac{2GM}{R} \rh n^t_{\pm} 
 + \frac{2mR}{GM} \left[ \lh 1 - \frac{GM}{R} \rh u^t - \ve \right] n^r_{\pm}, }\\
 \\
\dsp{ N^{r\vf}_{\pm} = - m \og_{\pm} R n^{\vf}_{\pm} - \frac{m}{R^2} \lh \eta + R^2 u^{\vf} \rh n^r_{\pm}, }\\
 \\ 
N^{t\vf}_{\pm} = \og^2_{\pm} ( \og_{\pm}^2 - \mu + \ag \kg + \gam \lb).
\ea
\label{ab.5}
\ee

\np
\bibliographystyle{unsrt}

\bibliography{vholten_relgeo_proceedings_2016}

\end{document}